\documentclass[11pt]{article}
\pdfoutput=1
\usepackage{jheppub,amsmath,amssymb}
\usepackage{bm}
\usepackage[table]{xcolor}
\usepackage{graphics}
\usepackage{tikz}
\usetikzlibrary{arrows,positioning,decorations.markings,decorations.pathmorphing,calc}
\usepackage{todonotes}

\definecolor{hyperref}{RGB}{026,028,087}

\def\gsim{ \lower .75ex \hbox{$\sim$} \llap{\raise .27ex \hbox{$>$}} }
\def\lsim{ \lower .75ex \hbox{$\sim$} \llap{\raise .27ex \hbox{$<$}} }
\def\be{\begin{equation}}
\def\ee{\end{equation}}
\def\bea{\begin{eqnarray}}
\def\eea{\end{eqnarray}}

\newcommand{\ba}{\begin{array}}
\newcommand{\ea}{\end{array}}
\newcommand{\mn}{\mu\nu}

\newcommand{\commentout}[1]{}

\newcommand{\comment}[1]{}

\newcommand{\bs}{\begin{split}}

\newcommand{\St}{{St\"uckelberg} }

\newcommand{\Ltd}{\mathcal{L}^\mathrm{TD}}
\newcommand{\T}{\mathrm{T}}

\def\ba{\begin{eqnarray}}
\def\ea{\end{eqnarray}}

\def\({\left(}
\def\){\right)}

\definecolor{jn}{RGB}{10, 10, 200} 
\definecolor{js}{RGB}{204, 0, 0} 
\definecolor{pgf}{RGB}{10, 150, 10} 

\newcommand*{\mathcolor}{}
\def\mathcolor#1#{\mathcoloraux{#1}}
\newcommand*{\mathcoloraux}[3]{%
  \protect\leavevmode
  \begingroup
    \color#1{#2}#3%
  \endgroup
}
\newlength{\stheight}
\newcommand\textst[1][fu-grey]{
	\ifmmode\setlength{\stheight}{+1.0ex}
	\else\setlength{\stheight}{+0.5ex}
	\fi
	\bgroup\markoverwith{\textcolor{#1}{\rule[\the\stheight]{2pt}{1.0pt}}}\ULon
} 

\newcommand{\textins}[2][fu-grey]{
	\ifmmode\mathcolor{#1}{#2}
	\else\textcolor{#1}{#2}\@\,
	\fi
}

\graphicspath{{./}}
\notoc

  \tikzstyle{vecArrow} = [thick, decoration={markings,mark=at position
   1 with {\arrow[semithick]{open triangle 60}}},
   double distance=1.4pt, shorten >= 5.5pt,
   preaction = {decorate},
   postaction = {draw,line width=1.4pt, white,shorten >= 4.5pt}]

\begin{document}

\title{Cycles of interactions in multi-gravity theories}

\author[a]{James H.C. Scargill}
\author[b]{, Johannes Noller}
\author[b]{, Pedro G. Ferreira}

\affiliation[a]{Theoretical Physics, University of Oxford, DWB, Keble Road, Oxford, OX1 3NP, UK} 
\affiliation[b]{Astrophysics, University of Oxford, DWB, Keble Road, Oxford, OX1 3RH, UK} 

\emailAdd{james.scargill@physics.ox.ac.uk}
\emailAdd{noller@physics.ox.ac.uk}
\emailAdd{p.ferreira1@physics.ox.ac.uk}

\abstract{
In this paper we study multi-gravity (multi-metric and multi-vielbein) theories in the presence of cycles of interactions (cycles in the so-called `theory graph'). It has been conjectured that in multi-metric theories such cycles lead to the introduction of a ghost-like instability, which, however, is absent in the multi-vielbein version of such theories. In this paper we answer this question in the affirmative by explicitly demonstrating the presence of the ghost in such multi-metric theories in the form of dangerous higher derivative terms in the decoupling limit Lagrangian; we also investigate the structure of interactions in the vielbein version of these theories and argue why the same ghost does not appear there. Finally we discuss the ramifications of our result on the dimensional deconstruction paradigm, which would seek an equivalence between such theories and a truncated Kaluza-Klein theory, and find that the impediment to taking the continuum limit due to a low strong-coupling scale is exacerbated by the presence of the ghost, when these theories are constructed using metrics.
}


\maketitle
\newpage

\section{Introduction} \label{sec-intro}

Theories of multiple interacting spin-2 fields have recently experienced a remarkable renaissance, though their history is long. It is known that theories of multiple interacting \emph{massless} spin-2 fields are inconsistent \cite{Boulanger:2000rq}, and thus a consistent theory of interacting spin-2 fields must necessarily involve a consistent theory of a massive spin-2 field.
Whilst a consistent linear theory of massive spin-2 was constructed in the 30's \cite{Fierz:1939ix}, for a long time it was thought that any non-linear extension would inevitably introduce the (in)famous Boulware-Deser ghost \cite{Boulware:1973my}, making the theory have an unacceptably low cutoff (generically $\Lambda_5 = (m^4 M_\text{Pl})^{1/5}$, where $m$ is the graviton mass), yet recently there was constructed \cite{deRham:2011qq, deRham:2011rn, deRham:2010kj, deRham:2010ik, Hassan:2011tf, Hassan:2011hr, Hassan:2012qv, Deser:2014hga} a theory which is ghost free, and has the higher cutoff of $\Lambda_3 = (m^2 M_\text{Pl})^{1/3}$. Furthermore it was shown that this theory retained its nice properties when extended to a theory of two dynamical spin-2 fields (bigravity) \cite{Hassan:2011ea, Hassan:2011zd}.
Work on further generalising this to a theory of an arbitrary number of interacting, dynamical spin-2 fields was then completed in \cite{Hinterbichler:2012cn}.
Some preliminary investigations into the cosmolgy of theories with more than two spin-2 fields was conducted in \cite{Tamanini:2013xia}.
Finally, ghosts aside, it has been argued that massive gravity may posses issues of acausality (for recent reviews discussing whether or not this is actually cause for concern see \cite{Deser:2014fta,deRham:2014zqa}); quite what ultimate bearing of this, especially on bi- and multi-gravity, is as yet unknown.

With more than two fields the possibility of constructing elaborate networks of interactions arises, and in \cite{Hinterbichler:2012cn} it was shown that \emph{provided the theory is formulated in terms of vielbeins}, any combination of individually healthy interactions would itself be healthy (though see \cite{Deffayet:2012nr, Deffayet:2012zc} for some questions about a hole in the proof). In \cite{Hinterbichler:2012cn} however it has been conjectured that the same is not true when the theory is formulated in terms of \emph{metrics}, and that if there is a cycle of interactions in the action, e.g. $A$ interacts with $B$, interacts with $C$, interacts with $A$ again, then the theory will again contain a ghost and cease to be healthy. This is the main question which we seek to address in the current paper.

Aside form the intrinsic theoretical interest in the question of which field theories are classically consistent, cycles of interacting spin-2 fields appear in another context: gravitational dimensional deconstruction \cite{ArkaniHamed:2001ca, ArkaniHamed:2003vb, Schwartz:2003vj, deRham:2013awa, Deffayet:2003zk}. This involves considering Einstein gravity on a discrete, periodic extra dimension, in order to compare it with the the Kaluza-Klein reduced version of the same theory, in which the infinite tower of states is truncated. The discretisation turns e.g. five dimensional GR into a four dimensional theory of multiple interacting spin-2 fields (different fields corresponding to different locations in the extra dimension), whilst its periodic nature (i.e. compactifying on $S^1$) means that the resulting theory will contain a cycle of interactions. Thus the question of whether such theories contain a ghost has bearing on the approach one must take to deconstructing gravitational dimensions.

This paper is structured in the following way: the next section briefly reviews some of the details of these theories, in particular their representation in terms of graphs, and their analysis via the \St trick. Section \ref{sec-plaquettes} then investigates a crucial way in which theories with cycles of interactions differ from purely tree-like interactions. That this difference will lead to ghosts in the metric version of the theory is then demonstrated in two different ways in section \ref{sec-ghosts}; in section \ref{sec-vielbein} we first review the vielbein version of multi-gravity theories, investigate the structure of interactions, and argue why the same ghost is not present there. Finally in section \ref{deconstructing dimensions} we discuss in more detail the link with dimensional deconstruction, before concluding in section \ref{sec-conc}.

\section{Interacting spin-2 fields and theory graphs} \label{sec-review}

Here we briefly review theories of multiple, interacting spin-2 fields, i.e. interacting massive gravitons, and in particular the \St analysis of such theories. For more detail see \cite{Noller:2013yja}.

\subsection{Theory graphs}

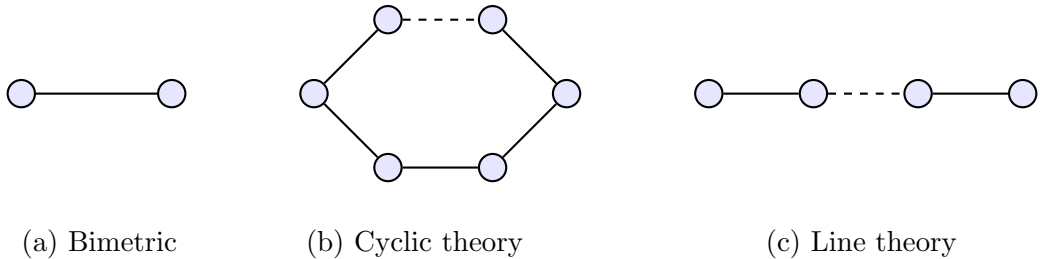
\begin{figure}[tp]
\centering
\begin{tikzpicture}[-,>=stealth',shorten >=0pt,auto,node distance=2cm,
  thick,main node/.style={circle,fill=blue!10,draw,font=\sffamily\large\bfseries},arrow line/.style={thick,-},barrow line/.style={thick,->},no node/.style={plain},rect node/.style={rectangle,fill=blue!10,draw,font=\sffamily\large\bfseries},red node/.style={rectangle,fill=red!10,draw,font=\sffamily\large\bfseries},green node/.style={circle,fill=green!20,draw,font=\sffamily\large\bfseries},yellow node/.style={rectangle,fill=yellow!20,draw,font=\sffamily\large\bfseries}]

 \node[main node](100){};
  \node[main node] (101) [right of=100] {};

\node[draw=none,fill=none](83)[right=0.7cm of 100]{};
    
   \node[main node] (1000)  [right=1.5cm of 101]{};
   \node[main node] (1001) [below right=1cm of 1000] {};
   \node[main node] (1002) [right=1cm of 1001] {};
   \node[main node] (1003) [above right=1cm of 1002] {};
   \node[main node] (1004) [above right=1cm of 1000] {};
   \node[main node] (1005) [above left=1cm of 1003] {};
   \node[draw=none,fill=none] (10000) [right=1cm of 1000] {};
   
  \node[main node] (2)  [right=1.5cm of 1003]{};
  \node[main node] (3) [right=1cm of 2] {};
   \node[main node] (4) [right=1cm of 3] {};
   \node[main node] (5) [right=1cm of 4] {};

\node[draw=none,fill=none](84)[right=0.3cm of 3]{};

  \path[every node/.style={font=\sffamily\small}]
  (100) edge node {} (101)
    (2) edge node {} (3)
     (4) edge node {} (5)
     (1000) edge node {} (1001)
     (1001) edge node {} (1002)
     (1002) edge node {} (1003)
     (1000) edge node {} (1004)
     (1003) edge node {} (1005);

\draw[-,dashed] (1004) to (1005);

\draw[-,dashed] (3) to (4);

   \node[draw=none,fill=none](92)[below of=83]{(a) Bimetric};
   
   \node[draw=none,fill=none](95)[below of=10000]{(b) Cyclic theory};
   
   \node[draw=none,fill=none](93)[below of=84]{(c) Line theory};
   
\end{tikzpicture}

\caption{Examples of different types of theory graphs: (a) isolated interactions simply connecting two fields not connected to any others; (b) a `cyclic theory' made up of $N$ sites with nearest neighbour interactions only, with the $N$-th site interacting with the first, hence forming a cycle with $N$ links; (c) a `line theory' made up of $N$ sites with nearest neighbour interactions only, forming a line with $N-1$ links} \label{fig-graph example}
\end{figure}

These theories can be represented using \emph{theory graphs} \cite{ArkaniHamed:2001ca,ArkaniHamed:2002sp,Hinterbichler:2012cn, Noller:2013yja,Noller:2014ioa} in which each field corresponds to a node of the graph; a term in the action which is an interaction between two fields corresponds to an edge of the graph, and an interaction between more than two fields can be represented by using an auxiliary vertex to which all the fields concerned are connected; see figure \ref{fig-graph example} for some examples. 

This formalism is useful as it allows one to restate certain questions about a particular theory in terms of properties of its theory graph, which is the main topic of this paper: looking at the effect of the presence of a cycle in the graph. We review previous work on this in the section \ref{cycles, dangerous}, but first mention another useful feature of theory graphs: they allow one to understand possible physical interpretations of the structure of a particular theory. In particular consider the `cyclic theory' depicted in figure \ref{fig-graph example}(b). Such a graph can be derived from discretising a circle, and hence one would expect that the theory resulting from the graph should be related to Kaluza-Klein reduction of the theory (in one more dimension) on a circle; this is known as `dimensional deconstruction' and we discuss it in more detail, and how it relates to the results of this paper, in section \ref{deconstructing dimensions}.

\subsection{Interacting spin-2 fields and the \St trick}

Each field will be dynamical and thus have a kinetic term in the action, which we take to be the Ricci scalar constructed out of that field; each field may then interact with one or more of the other fields. Since we are concerned in this paper with the question of the presence or absence of a ghost in certain theories, we take the interaction terms to be those which are known to be individually ghost-free: the dRGT interaction terms, which as interactions between two metrics $g$ and $f$ take the form
\be
\mathcal{S}_\text{int}[g,f] = \frac{1}{2} m^2 M^2_\text{Pl} \int \mathrm{d}^D x\, \sqrt{-\det g} \, e_m \left( \sqrt{ g^{-1} f } \right),
\ee
where $e_m(X)$ is the $m$-th order elementary symmetric polynomial in the eigenvalues of $X$. Rather than metrics the theory can also be formulated in terms of vielbeins (see \cite{Hinterbichler:2012cn, Ondo:2013wka, Gabadadze:2013ria} and references therein), in which case ghost-free interaction terms can be written which involve not just two, but up to $D$ different fields:
\be
\mathcal{S}_\text{int}[E_{(i_1)}, \dots, E_{(i_D)}] = \epsilon_{a_1 \dots a_D} \int E^{a_1}_{(i_1)} \wedge \dots \wedge E^{a_D}_{(i_D)}.
\ee
We will in fact only be concerned with interactions between at most two fields at a time, as these are the only ones for which ghost-free metric interaction terms can be written straightforwardly (see however \cite{Hassan:2012wt}). The equivalence between these two formulations breaks down in the presence of a cycle in the theory graph \cite{Hinterbichler:2012cn}.

\begin{figure}[htp]
\centering
\begin{tikzpicture}[->,>=stealth',shorten >=0pt,auto,node distance=2cm,
  thick,main node/.style={circle,fill=blue!10,draw,font=\sffamily\large\bfseries},arrow line/.style={thick,-},barrow line/.style={thick,->},no node/.style={plain}]
  
  \node[main node] (1) {i};
  \node[draw=none,fill=none] (3) [right=1cm of 1] {};
  \node[main node] (2) [right=1cm of 3] {j};

	\path[every node/.style={font=\sffamily\small}]
	(2) edge node [below] {$Y_{(i,j)}$} (1);
               
\end{tikzpicture}
\caption{The \St trick for an interaction term coupling $g_{(i)}$ and $g_{(j)}$.}
\label{fig-mapping}
\end{figure}
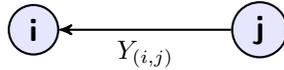

The kinetic terms individually respect a diffeomorphism invariance, 
\be
GC_i: \quad g_{(i)\mu\nu}(x) \to \partial_\mu f^\alpha \partial_\nu f^\beta g_{(i)\alpha\beta}(f(x)),
\ee
which can be written succinctly using functional composition notation
\be
GC_i: \quad g_{(i)} \to g_{(i)} \circ f.
\ee
So before the interaction terms are introduced a theory of $N$ fields respects $GC_1 \times \dots \times GC_N$; the interaction terms will (assuming that the theory graph is connected) break this down to the diagonal subgroup, in which every $GC$ acts in the same way. 
The full symmetry can be reintroduced via the \St trick: new (gauge) fields are introduced, mimicking the desired symmetry, which have just the right transformation properties to make the action invariant. 
We emphasise that the resulting (St\"uckelberg-ed) action is dynamically equivalent to the original action, the latter being a gauge-fixed version of the former.
As the kinetic terms are already invariant under the symmetries we are introducing, the \St fields will only enter via the interaction terms and it turns out there are several ways of doing this \cite{Noller:2013yja}. The approach we will consider throughout (with the exception of sections \ref{sec-w/o plaquettes} and \ref{sec-no plaquette vierbein}) is that in which each interaction term/link is considered separately; that is, for each interaction term one picks one field to be `mapped' onto the site of the other field. This is explained in table \ref{table-mapping}, and represented graphically in figure \ref{fig-mapping}.

\begin{table}[htp]
\centering
\begin{tabular}{ r | c c c }
& field & under $GC_i$: & under $GC_j$: \\
\hline
Before & $g_{(i)}$ & tensor & invariant \\
& $g_{(j)}$ & invariant & tensor \\
\hline
After & $g_{(i)}$ & tensor & invariant \\
& $G_{(i,j)} = g_{(j)} \circ Y_{(i,j)}$ & tensor & invariant \\
\end{tabular}
\caption{The \St trick for an interaction term coupling $g_{(i)}$ and $g_{(j)}$.}
\label{table-mapping}
\end{table}

Theories of multiple interacting \emph{massless} spin-2 fields are inconsistent \cite{Boulanger:2000rq}, and thus $N-1$ of the fields must be massive, however upon introduction of the \St fields all $N$ of the metrics obey a $GC$ gauge symmetry - the `lost' degrees of freedom are of course contained within the \St fields themselves. This is most easily seen by expanding each metric/vielbein about a flat background, and each \St field about the identity
\be
Y^\mu(x) = x^\mu + A^\mu,
\ee
followed by the introduction of an extra $U(1)$ symmetry: $A^\mu \to A^\mu + \partial^\mu \pi$. Upon taking the so-called `decoupling' limit
\be
m \to 0, \qquad M_\text{Pl} \to \infty, \qquad \Lambda_n = ( m^{n-1} M_\text{Pl} )^{\frac{1}{n}} \text{ fixed},
\ee
the $A^\mu$ fields transform as the helicity-1 components of the massive gravitons, $\pi$ as the helicity-0 components.

\subsection{Cycles and why they've been argued to be dangerous} \label{cycles, dangerous}

The possible importance of cycles in the theory graph when it comes to the ghost-freedom of a theory of multiple interacting spin-2 fields is first mentioned in \cite{Hinterbichler:2012cn}. The authors note that the equivalence between the vielbein and metric formulations of multi-gravity breaks down in the presence of a cycle, and whilst demonstrating the health of the vielbein theory go on to conjecture that the metric version will contain a Boulware-Deser ghost. The authors of \cite{Nomura:2012xr} (see also \cite{ASM-thesis}) then showed how the standard constraint analysis, which is used to prove the ghost-freedom of multi-metric theories with a tree-graph structure, breaks down in the presence of a cycle, again suggesting the presence of a ghost. For a related analysis in 3D see \cite{Afshar:2014dta}. 
Our paper now confirms this suspicion, by explicitly demonstrating the presence of higher derivative terms which will lead to a ghost, in the \St formulation.

\section{Plaquettes} \label{sec-plaquettes}

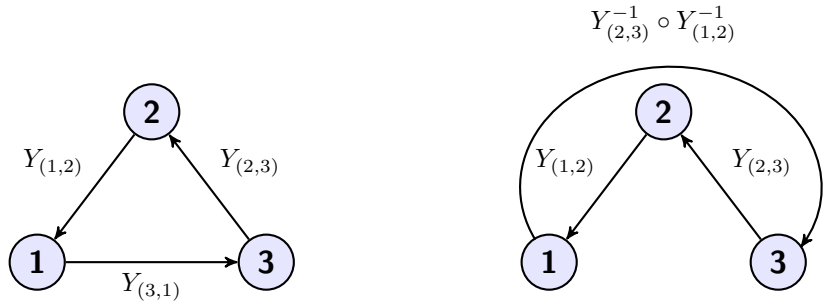
\begin{figure}[tp]
\centering
\begin{tikzpicture}[->,>=stealth',shorten >=0pt,auto,node distance=2cm,
  thick,main node/.style={circle,fill=blue!10,draw,font=\sffamily\large\bfseries},arrow line/.style={thick,-},barrow line/.style={thick,->},no node/.style={plain}]
  
  \node[main node] (1) {1};
  \node[draw=none,fill=none] (4) [right=1cm of 1] {};
  \node[main node] (3) [right=1cm of 4] {3};
  \node[main node] (2) [above of=4] {2};

	\path[every node/.style={font=\sffamily\small}]
	(2) edge node [above left] {$Y_{(1,2)}$} (1)
	(3) edge node [above right] {$Y_{(2,3)}$} (2)
	(1) edge node [below] {$Y_{(3,1)}$} (3);
  
  \node[main node] (11) [right=3cm of 3] {1};
  \node[draw=none,fill=none] (41) [right=1cm of 11] {};
  \node[main node] (31) [right=1cm of 41] {3};
  \node[main node] (21) [above of=41] {2};

	\path[every node/.style={font=\sffamily\small}]
	(21) edge node [above left] {$Y_{(1,2)}$} (11)
	(31) edge node [above right] {$Y_{(2,3)}$} (21);
         
         \node[draw=none,fill=none,style={font=\sffamily\small}] (81) [above=0.4 of 21] {$Y_{(2,3)}^{-1} \circ Y_{(1,2)}^{-1}$};
         
         \draw[<-] ($(31)+(0.3,0.2)$) arc(-30:206:2 and 1.6);
               
\end{tikzpicture}
\caption{Left: introducing a \St field for every link the case of a cycle leads to an overall constraint. Right: the constraint eliminates one \St field, replacing it with a plaquette formed from the other fields in the cycle.}
\label{plaquette figure}
\end{figure}

As noted in \cite{Noller:2013yja} one key difference between theory graphs with a cycle and those without (tree graphs) is that in the presence of a cycle there are now more links than broken copies of diffeomorphism invariance (since the diagonal subgroup remains unbroken). Hence if we introduce a \St field for every link as in the tree case, then we will end up with a set of fields which are not in fact independent, but satisfy some constraint. For example in the case of a trimetric cycle as depicted in figure \ref{plaquette figure}, the \St fields satisfy
\be
Y_{(1,2)} \circ Y_{(2,3)} \circ Y_{(3,1)} = \mathrm{id}. \label{constraint}
\ee

One way of dealing with this is to use the constraint to re-express one \St field in the cycle in terms of the others, i.e. in the trimetric case
\be
Y_{(3,1)} = Y_{(2,3)}^{-1} \circ Y_{(1,2)}^{-1}. \label{plaquette}
\ee
Following \cite{ArkaniHamed:2002sp} we call such a construction a \emph{plaquette}. Now site 1 is being pulled back to site 3 all the way around the cycle. This has the advantage that all the fields are now explicitly independent, however it does break the symmetry of the cycle by forcing one to pick a \St field to eliminate, as well as introducing interactions between \emph{all} the remaining \St fields through the plaquette. Thus one may wonder whether this is truly necessary, and in appendix \ref{necessity of plaquettes} we show that if one introduces more \St fields than broken copies of diffeomorphism invariance and treats them all independently, one encounters fields which are infinitely strongly coupled hindering the analysis.

\subsection{Plaquettes beyond the linear level}

The condition (\ref{plaquette}) yields for the \St scalars at linear order: $\pi_{(3,1)} = - (\pi_{(1,2)} + \pi_{(2,3)})$, however at higher order we can no longer look at the scalars and vectors separately. In fact
\begin{align}
Y^\mu_{(3,1)}(x) &= x^\mu + A^\mu_{(3,1)} + \partial^\mu \pi_{(3,1)} \\
= Y^{-1}_{(2,3)}\left(Y^{-1}_{(1,2)}(x) \right)^\mu &= x^\mu + \tilde{Z}^\mu_{(1,2)} + \tilde{Z}^\mu_{(2,3)} + \sum_{n=1}^\infty \frac{1}{n!} \tilde{Z}^{\nu_1}_{(1,2)} \dots \tilde{Z}^{\nu_n}_{(1,2)} \tilde{Z}^\mu_{(2,3), \nu_1 \dots \nu_n}, \label{Y-1Y-1}
\end{align}
where $Y^{-1,\mu}(x) = x^\mu + \tilde{Z}^\mu = x^\mu + B^\mu + \partial^\mu \phi$, so $B^\mu$ and $\phi$ are the dual fields associated with the \St vector $A^\mu$ and scalar $\pi$ (for more information see appendix \ref{dual fields}) and where a comma denotes partial differentiation. 
One can rewrite (\ref{Y-1Y-1}) as
\begin{align}
x^\mu &+ \partial^\mu \left( \phi_{(1,2)} + \phi_{(2,3)} + \sum_{n=1}^\infty \frac{1}{n!} \tilde{Z}_{(1,2)}^{\nu_1} \dots \tilde{Z}_{(1,2)}^{\nu_n} \phi_{(2,3), \nu_1 \dots \nu_n} \right) \\
&+ B_{(1,2)}^\mu + \sum_{n=0}^\infty \frac{1}{n!} \tilde{Z}_{(1,2)}^{\nu_1} \dots \tilde{Z}_{(1,2)}^{\nu_n} \left( B_{(2,3), \nu_1 \dots \nu_n}^\mu - \tilde{Z}_{(1,2)}^{\lambda,\mu} \phi_{(2,3), \lambda \nu_1 \dots \nu_n} \right),
\end{align}
from which the expressions for $A_{(3,1)}^\mu$ and $\pi_{(3,1)}$ can be read. And so we see that each receives contributions from both the vectors and the scalars, and in particular even if $A^\mu_{(1,2)}$ and $A^\mu_{(2,3)}$ are set to zero one still has $A^\mu_{(3,1)} \neq 0$. For example
\begin{align}
\pi_{(3,1)} =& - (\pi_{(1,2)} + \pi_{(2,3)} ) + \frac{1}{2} (\pi_{(1,2)} + \pi_{(2,3)} )^{,\mu} (\pi_{(1,2)} + \pi_{(2,3)} )_{,\mu} + A^\mu_{(1,2)} \pi_{(2,3),\mu} + \dots \\
A^\mu_{(3,1)} =& - A^\mu_{(1,2)} - A^\mu_{(2,3)} + A^\nu_{(1,2)} A^\mu_{(1,2),\nu} + A^\nu_{(2,3)} A^\mu_{(2,3),\nu} + A^\nu_{(1,2)} A^\mu_{(2,3),\nu} \nonumber \\
&+ (\pi_{(1,2)} + \pi_{(2,3)})^{,\nu} (A_{(1,2)}^\mu + A_{(2,3)}^\mu)_{,\nu} - \pi_{(2,3),\nu} ( A_{(1,2)}^{\mu,\nu} + A_{(1,2)}^{\nu,\mu}) \nonumber \\
&+ A_{(1,2)}^{\nu} \pi^{,\mu}_{(1,2),\nu} + A_{(2,3)}^\nu \pi^{,\mu}_{(2,3),\nu} - \pi_{(1,2)}^{,\mu\nu} \pi_{(2,3),\nu} + \dots. \label{A plaquette}
\end{align}
The final term of (\ref{A plaquette}) will turn out to have important consequences. It is also worth mentioning that there will be introduced quadratic mixing between the vectors through the kinetic term for the plaquette vector:
\be
\partial_{[\mu} A_{(3,1)\nu]} \partial^{[\mu} A^{\nu]}_{(3,1)} \supset \partial_{[\mu} A_{(1,2)\nu]} \partial^{[\mu} A^{\nu]}_{(1,2)} + 2\partial_{[\mu} A_{(1,2)\nu]} \partial^{[\mu} A^{\nu]}_{(2,3)} + \partial_{[\mu} A_{(2,3)\nu]} \partial^{[\mu} A^{\nu]}_{(2,3)}.
\ee
This is a qualitatively new feature, as in the absence of a plaquette only the tensors and scalars will be mixed quadratically (the scalar-tensor mixing can then be removed via a conformal transformation leaving just the scalars mixed).

\section{Ghosts in multi-metric theories} \label{sec-ghosts}

We will now show how the cycle leads to the introduction of a ghost at an energy scale \emph{below} $\Lambda_3$ (or equivalently leads to a lowering of the cutoff). For simplicity we consider a trimetric cycle, but in section \ref{sec-larger plaquettes} we consider larger cycles in the context of deconstructing dimensions; we also set all of the interactions strengths and Planck masses equal. For related work on trimetric cycle theories see \cite{Khosravi:2011zi,Nomura:2012xr}

The key point is that it is the $(1,2)$ and $(2,3)$ fields which are canonically normalised: $A_{(1,2)}^\mu \to \frac{1}{\Lambda_2^2} A_{(1,2)}^\mu$, $\pi_{(1,2)} \to \frac{1}{\Lambda_3^3} \pi_{(1,2)}$, and similarly for $(2,3)$. Thus the $(3,1)$ fields don't have the overall normalisation one would expect. In fact
\begin{align}
\pi_{(3,1)} &\to \sum_{n=0,m=1}^\infty \frac{1}{\Lambda_2^{2n} \Lambda_3^{3m}} A^n \pi^m,
\\
A^\mu_{(3,1)} &\to \sum_{n=1,m=0}^\infty \frac{1}{\Lambda_2^{2n} \Lambda_3^{3m}} A^n \pi^m +  \sum_{n=0,m=2}^\infty \frac{1}{\Lambda_2^{2n} \Lambda_3^{3m}} A^n \pi^m.
\end{align}
For the scalar this is not an issue, since $\Lambda_2 > \Lambda_3$, and so $\Lambda_2^{2n} \Lambda_3^{3m} \geq \Lambda_{3}^{2n+3m}$, thus any terms from the plaquette will sit at or above $\Lambda_3$. On the other hand $\Lambda_2^{2n} \Lambda_3^{3m} < \Lambda_{2}^{2n+3m}$ for $m > 0$, and so these terms from the vector will come in below $\Lambda_3$ (since $A^\mu \sim \frac{1}{\Lambda_2}$ is what is required to sit precisely at $\Lambda_3$).

More precisely, recalling that the interaction Lagrangian has an overall pre-factor $m^2 M_\text{Pl}^2$, we have
\begin{align}
m^2 M_\text{Pl}^2 \partial_{[\mu} A_{(3,1) \nu]} \partial^{[\mu} A_{(3,1)}^{\nu]} \supset &\frac{1}{\Lambda_4^4} \pi_{(1,2) ,\lambda [\mu} \pi_{(2,3) ,\nu]}^{,\lambda} \left( \partial^{[\mu} A_{(1,2)}^{\nu]} +\partial^{[\mu} A_{(2,3)}^{\nu]} \right) \nonumber \\ &-\frac{1}{\Lambda_4^8} \pi_{(1,2) ,\lambda [\mu} \pi_{(2,3) ,\nu]}^{,\lambda} \pi_{(1,2)}^{,\rho [\mu} \pi_{(2,3) ,\rho}^{,\nu]}, \label{ghost}
\end{align}
which we see to be higher derivative, but \emph{not} of such a form as to eliminate higher order equations of motion. Thus this theory contains a ghost associated with an energy scale $\Lambda_4 < \Lambda_3$.

There will be (an infinite number of) other, potentially dangerous, terms at energy scales between $\Lambda_4$ and $\Lambda_3$, however it is just the lowest energy scale which concerns us here, since this gives the new cutoff of the theory. Also one need not worry that this is just an artefact of some sort of truncation since (\ref{ghost}) are the \emph{only} terms at $\Lambda_4$.

Finally note that the first term in (\ref{ghost}) involves the \St vector \emph{linearly} - a qualitatively new feature, which means that it cannot classically be set to zero and ignored as it can in the absence of a cycle.

\subsection{Without plaquettes}\label{sec-w/o plaquettes}

The presence of these dangerous terms can also be demonstrated using a different method, in which one does not introduce $Y_{(3,1)}$ in the first place (and hence does not introduce a plaquette). This necessitates a slightly different approach to introducing the \St fields: one treats the action as a whole, picking one site onto which one maps all of the other fields \cite{Noller:2013yja}. For a trimetric cycle this means
\begin{align}
\mathcal{S}_\text{int} &= \mathcal{S}[g_{(1)}, g_{(2)}] + \mathcal{S}[g_{(2)}, g_{(3)}] + \mathcal{S}[g_{(3)}, g_{(1)}] \nonumber \\
&\to \mathcal{S}[g_{(1)} \circ Y_{(2,1)}, g_{(2)}] + \mathcal{S}[g_{(2)}, g_{(3)} \circ Y_{(2,3)}] + \mathcal{S}[g_{(3)} \circ Y_{(2,3)}, g_{(1)} \circ Y_{(2,1)}]; \label{w/o plaquette}
\end{align}
note the final term, which is different to all those considered previously, as it involves \St fields applied to both the metrics involved.

It turns out that for the pure scalar part of the action coming from this term one finds (e.g. for an interaction term consisting of just the first symmetric polynomial)
\begin{align}
&M_\text{Pl} \left[ (D-1) \mathcal{L}^\text{TD}_{(1)} \left( \pi_{(2,3)} \right) + \mathcal{L}^\text{TD}_{(1)} \left( \pi_{(2,1)} \right) \right]
+ \frac{1}{m^2} \left[\frac{1}{2} (D-2) \mathcal{L}^\text{TD}_{(2)} \left( \pi_{(2,3)} \right) + \mathcal{L}^\text{TD}_{(1,1)} \left( \pi_{(2,3)}, \pi_{(2,1)} \right) \right] \nonumber \\
&+ \frac{1}{\Lambda_5^5} \left[ \frac{1}{6} (D-3) \mathcal{L}^\text{TD}_{(3)} \left( \pi_{(2,3)} \right) + \frac{1}{2}\mathcal{L}^\text{TD}_{(2,1)} \left( \pi_{(2,3)}, \pi_{(2,1)} \right) \right] \nonumber \\
&+ \frac{1}{\Lambda_4^8} \left[ \frac{1}{24} (D-4) \mathcal{L}^\text{TD}_{(4)} \left( \pi_{(2,3)} \right) + \frac{1}{6}\mathcal{L}^\text{TD}_{(3,1)} \left( \pi_{(2,3)}, \pi_{(2,1)} \right) + \frac{1}{4} \pi_{(2,1) ,\lambda [\mu} \pi_{(2,3) ,\nu]}^{,\lambda} \pi_{(2,1)}^{,\rho [\mu} \pi_{(2,3) ,\rho}^{,\nu]} \right] + \dots 
\end{align}
where $\mathcal{L}^\text{TD}_{(n,l)}( \pi, \phi )$ is the total derivative combination of $n$ copies of $\partial^2 \pi$ and $l$ of $\partial^2 \phi$. We see that this takes the expected, safe form, i.e. a total derivative, at quadratic and cubic order, but at quartic order a new type of term appears which is precisely the same\footnote{Recall that $Y_{(2,1)} = Y_{(1,2)}^{-1}$, and so $\pi_{(2,1)} = - \pi_{(1,2)} + \dots$} as that in (\ref{ghost}) suppressed by $\Lambda_4^8$. Similarly the vector-scalar-scalar terms will consist of total derivatives along with the term from (\ref{ghost}) suppressed by $\Lambda_4^4$.

Of course this is to be expected, as the different ways of introducing the \St fields are all equivalent \cite{Noller:2013yja}; in fact, performing a gauge transformation on the final term (each term is gauge invariant, so they can be treated individually) in (\ref{w/o plaquette}) with parameter $Y^{-1}_{(2,3)}$, and noting that $Y_{(2,1)} = Y_{(1,2)}^{-1}$ one has
\be
\mathcal{S}\left[g_{(3)}, g_{(1)} \circ \left( Y_{(1,2)}^{-1} \circ Y_{(2,3)}^{-1} \right) \right],
\ee
which is identical to using a plaquette.

\section{Absence of ghost in multi-vielbein theories} \label{sec-vielbein}

We show below how the dangerous terms which arise in multi-metric theories do not do so in the case of multi-vielbein theories, which is to be expected since such theories have been shown to be ghost-free even when cycles are present in the theory graph \cite{Hinterbichler:2012cn}. For simplicity we specialise to $D = 4$, and where noted, interaction terms which consist of just the first symmetric polynomial, however the result is completely general.

\subsection{Vierbein version of bi-gravity}

First we recapitulate the vierbein version of bi-gravity, including how to apply the \St trick and demonstrating the equivalence to the metric version, which breaks down in the presence of cycles. \cite{Hinterbichler:2012cn, Hassan:2012wt, Deffayet:2012nr, Ondo:2013wka, Gabadadze:2013ria}

The Einstein-Hilbert action becomes
\be
\frac{M_{\text pl}^2}{2} \int \mathrm{d}^4x\,\sqrt{-g} \, R \to \frac{M_{\text{Pl}}^2}{2} \int \epsilon_{abcd} E^a \wedge E^b \wedge R^{cd}(E),
\ee
where $E^a$ is a one-form vierbein and $R^{ab}(E)$ is the associated gauge curvature two-form, and the interaction terms become
\be
\frac{m^2 M_{\text{Pl}}^2}{4} \int \mathrm{d}^4x\,\sqrt{-g} \, e_m \left( \sqrt{g^{-1} f} \right) \to \frac{m^2 M_{\text{Pl}}^2}{4} \int \epsilon_{a_1 \dots a_{4-m} b_1 \dots b_m} E^{a_1} \wedge \dots \wedge E^{a_{4-m}} \wedge F^{b_1} \wedge \dots \wedge F^{b_m},
\ee
where $F^a$ is a second one-form vierbein (distinct from $E^a$).
Each Einstein-Hilbert term now manifestly respects both a copy of diffeomorphism invariance, $E^a_\mu(x) \to E^a_\nu(f(x)) \partial_\mu f^\nu$, and also of local Lorentz invariance, $E^a \to \Lambda^a_{\phantom{1}b} E^b$, which like the diff invariance is broken down to a single (diagonal) copy by the interaction terms. Thus when applying the \St trick, it makes sense to not only introduce diff \St fields, but also ones to reintroduce the local Lorentz invariances (these \St fields will conventionally be denoted by $\Lambda$).

It is clear from the form of the symmetry that the Lorentz \St field will be non-dynamical and its equation of motion yields 
\be
E^a_{[\mu|}\eta_{ab} (\Lambda F)^b_{|\nu]} = 0, \label{DvN}
\ee
for any combination of interaction terms. In unitary gauge ($\Lambda^a_{\phantom{1}b} = \delta^a_b$) this becomes the famous Deser-van-Niewenhuizen (DvN) symmetric vierbein condition. We will now show how this condition is sufficient to show the equivalence with the metric version of bi-gravity (in four dimensions it is also necessary \cite{Deffayet:2012zc}). In matrix notation (\ref{DvN}) reads $E^\T \eta (\Lambda F) = (\Lambda F)^\T \eta E$ from which we get $(\Lambda F) E^{-1} = \eta^{-1} (E^\T)^{-1} (\Lambda F)^\T \eta$, and thus
\be
(E^{-1} \Lambda F)(E^{-1} \Lambda F) = E^{-1} \eta^{-1} (E^{-1})^\T (\Lambda F)^\T \eta (\Lambda F) = (E^\T \eta E)^{-1} (F^\T \eta F) = g^{-1} f,
\ee
where the metric $g_{\mn} = E^a_\mu \eta_{ab} E^b_\nu$, and similarly for $f$ and $F$. Therefore $E^{-1} (\Lambda F) = \sqrt{g^{-1} f}$ (modulo non-uniqueness of the square root), and hence
\begin{align}
\int &\epsilon_{a_1 \dots a_{4-m} b_1 \dots b_m} E^{a_1} \wedge \dots \wedge E^{a_{4-m}} \wedge (\Lambda F)^{b_1} \wedge \dots \wedge (\Lambda F)^{b_m} \nonumber \\
&= \int d^4x \, \epsilon_{a_1 \dots a_{4-m} b_1 \dots b_m} \epsilon^{\mu_1 \dots \mu_{4-m} \nu_1 \dots \nu_m} E^{a_1}_{\mu_1} \dots E^{a_{4-m}}_{\mu_{4-m}} (\Lambda F)^{b_1}_{\nu_1} \dots (\Lambda F)^{b_m}_{\nu_m} \nonumber \\
&= \int d^4x \, |E|\, \epsilon_{\mu_1 \dots \mu_{4-m} \rho_1 \dots \rho_m} \epsilon^{\mu_1 \dots \mu_{4-m} \nu_1 \dots \nu_m} (E^{-1})^{\rho_1}_{b_1}(\Lambda F)^{b_1}_{\nu_1} \dots (E^{-1})^{\rho_m}_{b_m}(\Lambda F)^{b_m}_{\nu_m} \nonumber \\
&= \int d^4x \, \sqrt{-g}\, \frac{1}{m!} \delta_{\rho_1 \dots \rho_m}^{\nu_1 \dots \nu_m} \sqrt{g^{-1}f}^{\rho_1}_{\nu_1} \dots \sqrt{g^{-1}f}^{\rho_m}_{\nu_m} = \int d^4x \, \sqrt{-g}\, e_m \left( \sqrt{g^{-1} f} \right), \nonumber
\end{align}
where (in 4D) we have defined a tensor $\delta^{\alpha_1 \ldots \alpha_n}_{\beta_1 \ldots \beta_n}$ separately anti-symmetric in its indices $\alpha_1 \ldots \alpha_n$ and $\beta_1 \ldots \beta_n$ in terms of the totally antisymmetric tensor $\varepsilon$ via
\be \label{delta-def}
\delta^{\alpha_1 \ldots \alpha_n}_{\beta_1 \ldots \beta_n}  \equiv \frac{1}{(4-n)!}\varepsilon^{\alpha_1 \ldots \alpha_n \lambda_1 \ldots \lambda_{4-n}}
\varepsilon_{\beta_1 \ldots \beta_n \lambda_1 \ldots \lambda_{4-n}}.
\ee

It is clear that in the case of multi-gravity, \emph{in the absence of cycles in the theory graph}, this equivalence of the vierbein and metric versions will continue to hold, since each Lorentz \St field is independent and hence each pair of vierbeine joined by an interaction term will individually obey the DvN condition (\ref{DvN}).

\subsection{Decoupling Limit}

Just as in the metric version one can then perturb about a flat background for the vierbeine, 
\be
E^a_\mu = \delta^a_\mu + \frac{1}{2 M_\text{Pl}} h^a_\mu, \qquad F^a_\mu = \delta^a_\mu + \frac{1}{2 M_\text{Pl}} l^a_\mu,
\ee
and about the identity for the \St fields,
\be
\partial_\mu Y^\nu = \delta^\nu_\mu + \frac{1}{m M_\text{Pl}} \partial_\mu A^\nu + \Pi^\nu_\mu, \qquad \Lambda^a_{\phantom{1}b} = e^{\frac{1}{m M_\text{Pl}} \omega^a_{\phantom{1}b}},
\ee
where the fields have already been canonically normalised, and $\Pi^\mu_\nu = \frac{1}{m^2 M_\text{Pl}}\pi^{,\mu}_{,\nu}$; the decoupling limit is then taken in the usual way:
\be
M_\text{Pl} \to \infty, \qquad m \to 0, \qquad \Lambda_3 = (m^2 M_\text{Pl} )^\frac{1}{3} \quad \text{fixed}.
\ee
The normalisation of $\omega$ may seem arbitrary, since it has no kinetic term, however due to its antisymmetry, $\omega$ will only couple to $\partial_\mu A^\nu$ at leading order, and hence it must have the same scaling in order to survive the decoupling limit without generating any divergent terms.

Since it scales in the same way as $\partial_\mu A^\nu$, we know that no terms involving both $\omega$ and a helicity-2 field will survive the decoupling limit. Therefore the helicity-2/0 part of the action will be of exactly the same form as in the metric version. For the helicity-1/0 part one finds (simplifying to the case of a single interaction term)
\begin{align}
\mathcal{S}_\text{int} &= - \frac{m^2 M^2_\text{Pl}}{2} \int \frac{1}{3!} \epsilon_{abcd} E^a \wedge E^b \wedge E^c \wedge F^d \\
\to \mathcal{S}_{1/0} &= -\frac{1}{4} \int d^4x \, \left( \delta^{\mu\nu}_{ab} G^a_{\phantom{1}\mu} \omega^b_{\phantom{1}\nu} + \delta^{\mu\nu}_{ab} (1 + \Pi)^a_\mu \omega^b_{\phantom{1}\lambda} \omega^\lambda_{\phantom{1}\nu} + \delta^{\mu\nu\lambda}_{abc} (1 + \Pi)^a_\mu \omega^b_{\phantom{1}\nu} \omega^c_{\phantom{1}\mu} \right) \\
&= \frac{1}{4} \int d^4x \, \left( [ G \omega ] - [ (1 + \Pi) \omega^2 ] \right), \label{S10 vierbein line}
\end{align}
where $G_{\mu\nu} = 2 \partial_{[ \mu} A_{\nu ]}$, and $[ M ] = \mathrm{tr} M$.

The Lorentz \St field is an auxiliary field, whose equation of motion (\ref{DvN}) in the decoupling limit becomes
\be
G_{\mu\nu} = 2 ( \omega_{\mu\nu} + \omega_{[\mu | \lambda} \Pi^\lambda_{| \nu ]} ) \label{LS eom line}.
\ee
As a matrix equation this is the Lyapunov equation, which has solution
\be
\omega_{\mu\nu} = \int_{0}^{\infty} du \, e^{-2u} e^{-u \Pi^\rho_\mu} G_{\rho \lambda} e^{-u \Pi^\lambda_\nu} = \sum_{n,m = 0}^{\infty} \frac{(-1)^{n+m}}{2^{1+n+m}} {}^{n+m}C_n \left( \Pi^n G \Pi^m \right)_{\mu\nu},
\ee
and upon substitution of this into (\ref{S10 vierbein line}) we find
\be
\mathcal{S}_{1/0} = \frac{1}{4} \int d^4x \, \left( - \frac{1}{4} \left[ G^2 \right] + \sum_{n=1}^\infty \frac{(-1)^n}{2^{2+n}} \sum_{m=0}^n \left( (n-1) {}^{n-1}C_{m-1} - (n+1) {}^{n-1}C_m \right) \left[ \Pi^{n-m} G \Pi^m G \right] \right). \label{S10 single link final}
\ee

The equivalent calculation in the metric version is slightly more involved but we have confirmed that it nonetheless yields the same result, as it should given the equivalence, for bigravity, demonstrated in the previous section.

\subsection{A simple cycle}

Let us now see what the vierbein version of a trimetric cycle looks like. The fact that not all three Lorentz \St fields are independent means that instead of (\ref{DvN}) for each pair of vierbeins we now have
\begin{align}
| E_{(1)} | E^a_{(1),[\mu|} \eta_{ab} (\Lambda_{(1,2)} E_{(2)})^b_{|\nu]} - | E_{(2)} | E^a_{(2),[\mu|} \eta_{ab} (\Lambda_{(2,3)} E_{(3)})^b_{|\nu]} &= 0, \\
| E_{(2)} | E^a_{(2),[\mu|} \eta_{ab} (\Lambda_{(2,3)} E_{(3)})^b_{|\nu]} - | E_{(3)} | E^a_{(3),[\mu|} \eta_{ab} (\Lambda_{(3,1)} E_{(1)})^b_{|\nu]} &= 0,
\end{align}
and we no longer have direct equivalence with the metric version. 
The dangerous terms found in section \ref{sec-ghosts} did not involve the helicity-2 mode, so let us focus on the helicity-1/0 part:
\begin{align}
-4 \mathcal{L}_{1/0} = &\sum_{i=1}^2 \left[ \omega_{(i,i+1)\mu\nu} G^{\mu\nu}_{(i,i+1)} + \left(1 + \Pi_{(i,i+1)} \right)^\mu_\nu \omega^\nu_{(i,i+1)\rho} \omega^\rho_{(i,i+1)\mu} \right] \nonumber \\
&+ m^2 M_\text{Pl}^2 \left( \omega_{(3,1)\mu\nu} G^{\mu\nu}_{(3,1)} + (1 + \Pi_{(3,1)} )^\mu_\nu \omega^\nu_{(3,1)\rho} \omega^\rho_{(3,1)\mu} \right) + \mathcal{O}(\omega^3). \label{vierbein pre plaq}
\end{align}
As in the metric case (\ref{Y-1Y-1}) can be used to replace the diff \St fields, which we will write as
\begin{align}
A^\mu_{(3,1)} &= \sum_{n=2}^\infty \frac{1}{\Lambda_3^{3n}} a_n^\mu + \frac{1}{\Lambda_2^2} \sum_{n=0}^\infty \frac{1}{\Lambda_3^{3n}} b_n^\mu + \mathcal{O}\left(\frac{1}{\Lambda_2^4} \right), \\
\pi_{(3,1)} &= \sum_{n=1}^\infty \frac{1}{\Lambda_3^{3n}} \sigma_n + \mathcal{O}\left(\frac{1}{\Lambda_2^2} \right).
\end{align}
Similarly the Lorentz \St field $\omega_{(3,1)}$ can be related to the others via
\be
 e^{\omega_{(3,1)}} = \Lambda_{(3,1)} = \Lambda_{(2,3)}^{-1}\Lambda_{(1,2)}^{-1} = e^{-\omega_{(2,3)}}e^{-\omega_{(1,2)}} = e^{-\frac{1}{\Lambda_2^2}(\omega_{(1,2)} + \omega_{(2,3)}) - \frac{1}{2\Lambda_2^4} [\omega_{(1,2)} , \omega_{(2,3)} ] + \dots },
 \ee
where we have chosen to normalise $\omega_{(1,2)}$ and $\omega_{(2,3)}$ by $\Lambda_2^2$. (\ref{vierbein pre plaq}) then becomes
\begin{align}
-4 \mathcal{L}_{1/0} = &\frac{1}{2} \omega_{+ \mu\nu} \left\{ G_+^{\mu\nu} - 4\partial^{[\mu}(b + \Lambda_2^2 a)^{\nu]} \right\} + \frac{1}{2} \omega_{- \mu\nu} G_-^{\mu\nu} \nonumber \\
 &+ \left( 1 + \partial^\mu_\nu \sigma + \partial^\mu a_\nu \right) \omega^\nu_{+ \rho} \omega^\rho_{+ \mu} + \frac{1}{4} \Big\{ (2 + \Pi_+ )^\mu_\nu \left( \omega^\nu_{+ \rho} \omega^\rho_{+ \mu} + \omega^\nu_{- \rho} \omega^\rho_{- \mu} \right) \nonumber \\
 &+ ( \Pi_-^{\mu\nu} + \partial^{[\mu} a^{\nu]} ) \omega_{+ \nu \rho} \omega^\rho_{- \mu} + ( \Pi_-^{\mu\nu} - \partial^{[\mu} a^{\nu]} ) \omega_{- \nu \rho} \omega^\rho_{+ \mu} \Big\} + \mathcal{O}\left(\frac{1}{\Lambda_2^2} \right), \label{vielbein plaquette lagrangian}
\end{align}
where $\omega_{\pm} = \omega_{(1,2)} \pm \omega_{(2,3)}$, etc. The terms shown are those which naively would survive the decoupling limit holding $\Lambda_3$ constant (\emph{which we do not yet take}). One derives the following equations of motion for the Lorentz \St fields:
\begin{align}
\big( G_+ - 4 \partial (b + \Lambda_2^2 a) + \omega_+ (6 + \Pi_+ + 4 \partial^2 \sigma + 2(\partial a + (\partial a)^T) ) - & \omega_- (\Pi_- + (\partial a - (\partial a)^T ) ) \big)^{[\mu\nu]} \nonumber \\
+ \mathcal{O}\left(\frac{1}{\Lambda_2^2} \right) &= 0, \label{Lorentz Stueckelberg eom 1} \\
\left( G_- + \omega_- (2 + \Pi_+) - \omega_+ (\Pi_- - (\partial a - (\partial a)^T ) ) \right)^{[\mu\nu]} + \mathcal{O}\left(\frac{1}{\Lambda_2^2} \right) &= 0,
\end{align}
and can attempt to solve them via an expansion in powers of $\Lambda_2$ and $\Lambda_3$. Doing so one finds that the leading terms are
\be
\omega_+^{\mu\nu} = \frac{4}{3}\frac{\Lambda_2^2}{\Lambda_3^6} \partial^{[\mu} a_2^{\nu]} + \dots, \qquad \text{and} \qquad \omega_-^{\mu\nu} = \frac{4}{3}\frac{\Lambda_2^2}{\Lambda_3^6} \Pi_{-\lambda}^{[\mu} \partial^{\lambda} a_2^{\nu]} + \dots. \label{leading terms vielbein plaquette}
\ee
But we immediately see a problem: with these solutions, terms in (\ref{vielbein plaquette lagrangian}) which we have ignored in fact will contribute at a level equivalent to those we have kept. Or in other words, $\omega$ should not be normalised by $\Lambda_2$, but by $\Lambda_3$, and so if we want to take the decoupling limit keeping $\Lambda_3$ fixed, we must include terms with arbitrary powers of $\omega$. Whilst this does not mean that taking such a decoupling limit is impossible, it certainly complicates matters, to the extent that unfortunately we are unable to explicitly show the absence of the ghost in this way.

\subsubsection{Without a plaquette} \label{sec-no plaquette vierbein}

We can of course analyse the trimetric cycle in the same manner as section \ref{sec-w/o plaquettes} - pulling everything back to one site. 
The parts of the interaction Lagrangian involving just one \St field, i.e. $\mathcal{L}[E_{(1)} \circ Y_{(2,1)} , E_{(2)} ] + \mathcal{L}[ E_{(2)} , E_{(3)} \circ Y_{(2,3)} ]$, will have standard forms and so we just need to consider the part involving two \St fields:
\begin{align}
&\mathcal{L}[E_{(3)} \circ Y_{(2,3)} , E_{(1)} \circ Y_{(2,1)} ] = \nonumber \\
&-\frac{m^2 M_\text{Pl}^2}{2}\frac{1}{3!} \delta^{\mu\nu\rho\sigma}_{abcd} (\Lambda_{(2,3)} E_{(3)} \partial Y_{(2,3)})^a_\mu (\Lambda_{(2,3)} E_{(3)} \partial Y_{(2,3)})^b_\nu (\Lambda_{(2,3)} E_{(3)} \partial Y_{(2,3)})^c_\rho (\Lambda_{(2,1)} E_{(1)} \partial Y_{(2,1)})^d_\sigma. \label{double St link vierbein}
\end{align}
Expanding around a flat background and normalising the fields in the usual way (\ref{double St link vierbein}) becomes
\begin{align}
-4\mathcal{L} =  &\Lambda_3^3 \left( \frac{1}{3} h_{(1) \mu\nu} \tilde{X}^{\mu\nu}_{(0,3)} + h_{(3) \mu\nu} \tilde{X}^{\mu\nu}_{(1,2)} \right) \nonumber \\
&+ \frac{1}{3} \left(\omega_{(2,1)\mu\lambda} \partial_\nu A^\lambda_{(2,1)} + \frac{1}{2} \left( 1 + \Pi_{(2,1)} \right)_\mu^\lambda \omega_{(2,1)\lambda\rho} \omega_{(2,1)\nu}^\rho \right) \tilde{X}^{\mu\nu}_{(0,3)} \nonumber \\
&+  \left(\omega_{(2,3)\mu\lambda} \partial_\nu A^\lambda_{(2,3)} + \frac{1}{2} \left( 1 + \Pi_{(2,3)} \right)_\mu^\lambda \omega_{(2,3)\lambda\rho} \omega_{(2,3)\nu}^\rho \right) \tilde{X}^{\mu\nu}_{(1,2)} \nonumber \\
&+ (\omega_{(2,1)\mu\nu} + \partial_\nu A_{(2,1)\mu}) (\omega_{(2,1)\rho\sigma} + \partial_\sigma A_{(2,1)\rho}) \tilde{X}^{\mu\nu\rho\sigma}_{(0,2)} \nonumber \\
&+ 2(\omega_{(2,3)\mu\nu} + \partial_\nu A_{(2,3)\mu}) (\omega_{(2,3)\rho\sigma} + \partial_\sigma A_{(2,3)\rho}) \tilde{X}^{\mu\nu\rho\sigma}_{(1,1)} \nonumber \\
&+ \Lambda_2^2 \left( \frac{1}{3} (\omega_{(2,1)\mu\nu} + \partial_\nu A_{(2,1)\mu}) \tilde{X}^{\mu\nu}_{(0,3)} + (\omega_{(2,3)\mu\nu} + \partial_\nu A_{(2,3)\mu}) \tilde{X}^{\mu\nu}_{(1,2)} \right) \nonumber \\
&+ \Lambda_2^2 \left( \frac{1}{3} \omega_{(2,1)\mu\nu} \Pi^\nu_{(2,1)\lambda} \tilde{X}^{\mu\lambda}_{(0,3)} + \omega_{(2,3)\mu\nu} \Pi^\nu_{(2,3)\lambda} \tilde{X}^{\mu\lambda}_{(1,2)} \right) + \mathcal{O}\left( \frac{1}{\Lambda_2^2} \right),
\end{align}
where $\tilde{X}^{\mu\nu \dots \rho\sigma}_{(n,m)} = \eta^{\nu\tilde{\nu}} \dots \eta^{\sigma\tilde{\sigma}} \delta^{\mu \dots \rho \alpha_1 \dots \alpha_n \gamma_1 \dots \gamma_m}_{\tilde{\nu} \dots \tilde{\sigma} \beta_1 \dots \beta_n \delta_1 \dots \delta_m}  (1 + \Pi_{(2,1)})_{\alpha_1}^{\beta_1} \dots
(1 + \Pi_{(2,3)})_{\gamma_1}^{\delta_1} \dots
$. 
The terms on the final two lines would lower the cutoff since they are suppressed by a scale below $\Lambda_3$; we see that those in the penultimate line do not contribute because $\omega_{\mu\nu} \tilde{X}^{\mu\nu} = 0$ since $\tilde{X}$ is symmetric whereas $\omega$ is antisymmetric, and $\partial_\nu A_\mu \tilde{X}^{\mu\nu} = \partial_\nu ( A_\mu \tilde{X}^{\mu\nu} )$ since $\partial_\mu \tilde{X}^{\mu\nu} = 0$; the terms in the final line however do not disappear\footnote{We thank Garrett Goon and Kurt Hinterbichler for alerting us to this.} and, as we show below, one arrives at a similar conclusion to the previous section.

Including the contributions from the other links, leads to 
\begin{align}
\mathcal{L} \supset &-\omega_{(2,1)\mu\nu} \left( \left( 3 \tilde{X}^{\mu\lambda}_{(2,0)} + \frac{1}{3} \tilde{X}^{\mu\lambda}_{(0,3)} \right) G_{(2,1) \lambda}^{\nu} + \left( 8 \tilde{X}^{\mu\nu\lambda\rho}_{(1,0)} + 2 \tilde{X}^{\mu\nu\lambda\rho}_{(0,2)} \right) G_{(2,1) \lambda\rho} + \frac{\Lambda_2^2 }{3} \Pi^\nu_{(2,1)\lambda} \tilde{X}^{\mu\lambda}_{(0,3)} \right) \nonumber \\
&- \omega_{(2,3)\mu\nu} \left( \left( \tilde{X}^{\mu\lambda}_{(1,2)} - \eta^{\mu\lambda} \right) G_{(2,3) \lambda}^{\nu} + 4 \tilde{X}^{\mu\nu\lambda\rho}_{(1,1)} G_{(2,3) \lambda\rho} + \Lambda_2^2 \Pi^\nu_{(2,3)\lambda} \tilde{X}^{\mu\lambda}_{(1,2)} \right) \nonumber \\
&+ \omega_{(2,1)\mu\nu} \omega_{(2,1)\lambda\rho} \left( \left( \frac{3}{2} \tilde{X}^{\mu\sigma}_{(2,0)} + \frac{1}{6} \tilde{X}^{\mu\sigma}_{(0,3)} \right) (1 + \Pi_{(2,1)} )^\rho_\sigma \eta^{\nu\lambda} + 4 \tilde{X}^{\mu\nu\lambda\rho}_{(1,0)} + \tilde{X}^{\mu\nu\lambda\rho}_{(0,2)} \right) \nonumber \\
&+ \omega_{(2,3)\mu\nu} \omega_{(2,3)\lambda\rho} \left( \left( \frac{1}{2} \tilde{X}^{\mu\sigma}_{(1,2)} (1 + \Pi_{(2,3)} )^\rho_\sigma + \Pi_{(2,3)}^{\mu\rho} - \eta^{\mu\rho} \right) \eta^{\nu\lambda} + 2 \tilde{X}^{\mu\nu\lambda\rho}_{(1,1)} \right) + \mathcal{O}\left( \frac{1}{\Lambda_2^2} \right),
\end{align}
from which we derive the following equations of motion for the Lorentz \St fields:
\be
T_i^{\mu\nu} = \omega_{(2,i)}^{\mu\nu} - 2 \left( \omega_{(2,i)\lambda}^{[\mu} C_i^{\nu]\lambda} + A_i^{\lambda [\mu} \omega_{(2,i) \lambda \rho} B_i^{\nu] \rho} \right) + \mathcal{O}\left( \frac{1}{\Lambda_2^2} \right), \label{loop lst eom}
\ee
where $i = 1,3$ and $T_i$ etc. are given in appendix \ref{loop lst eom terms}.
Compared to  (\ref{LS eom line}), the equivalent for a single link, equation (\ref{loop lst eom}) is more complicated and for completeness its solution neglecting the terms which are naively suppressed by $\Lambda_2^{2}$ is given in appendix \ref{loop lst eom solve}. For now we only need look at the terms suppressed by the lowest scale, for which we find
\be
\omega_{(2,1)}^{\mu\nu} = 2 \frac{\Lambda_2^2}{\Lambda_3^6} \Pi^{[\mu}_{(2,1)\lambda} \Pi^{\nu] \lambda}_{(2,3)} + \dots, \qquad \omega_{(2,3)}^{\mu\nu} = - 6 \frac{\Lambda_2^2}{\Lambda_3^6} \Pi^{[\mu}_{(2,1)\lambda} \Pi^{\nu] \lambda}_{(2,3)} + \dots,
\ee
where we have explicitly extracted $\Lambda_3^{-3}$ from each $\Pi$. These exhibit the same scaling as (\ref{leading terms vielbein plaquette}) in the previous section. Our conclusion is thus the same: in order to consistently take the decoupling limit holding $\Lambda_3$ fixed one must consider terms with an arbitrary nuber of $\omega$'s.

In the absence of an explicit re-summation of the $\omega$-dependent contributions we cannot prove ghost-freedom in this way, but the fact that this is different from the metric version, and the results of \cite{Hinterbichler:2012cn} (though see  \cite{Deffayet:2012nr, Deffayet:2012zc} for some questions about a hole in the proof) inspire confidence in the ghost-freedom of the vielbein version.

\section{Cycles and deconstructing dimensions} \label{deconstructing dimensions}

Dimensional deconstruction \cite{ArkaniHamed:2001ca, ArkaniHamed:2002sp, ArkaniHamed:2003vb, Schwartz:2003vj, deRham:2013awa} is the idea that a theory placed on a discrete, periodic extra dimension is equivalent to the truncation of the infinite tower of modes which arises from a standard KK reduction on $S^1$. In this way it allows one to consider whether a low energy effective theory can be derived from the compactification of a higher dimensional theory. 

Such a discrete, periodic extra dimension can clearly be represented as a circle theory graph \cite{ArkaniHamed:2003vb,Schwartz:2003vj}, as in figure \ref{fig-graph example}(b), and thus analysis of the dimensional deconstruction paradigm requires analysis of theory graphs containing cycles. In particular, as we will want to take the $N \to \infty$ naive continuum limit we will need to consider larger cycles than in the previous sections.

\subsection{Larger plaquettes} \label{sec-larger plaquettes}

Whilst in the trimetric case it is possible to avoid the use of plaquettes, simplifying matters slightly, for larger cycles the use of plaquettes (or plaquette-like constructions) is unavoidable, since now not every site is one link removed from every other site. Thus we now look at plaquettes of larger size.

In the case of a cycle of $N$ metrics the plaquette expression (\ref{plaquette}) is extended in the obvious way and the equivalent of the final term in (\ref{A plaquette}) is
\be
A^\mu_{(N,1)} \supset -\sum_{i=1}^{N-2} \sum_{j=i+1}^{N-1} \pi^{,\mu}_{(i,i+1),\lambda} \pi^{,\lambda}_{(j,j+1)} \label{larger plaquette A}
\ee
and the analysis proceeds in the same way as in the trimetric case, leading to ghost-inducing terms of the same form as (\ref{ghost}), except that now there are $\frac{1}{2}(N-1)(N-2)$ vector-scalar-scalar terms\footnote{One might expect an additional factor of $N-1$ from $\sum_i \partial A_i$, however the vector modes must be demixed and this will be precisely one of the propagating modes.} and $\frac{1}{8}(N-1)(N-2)(N^2 -3N +4)$ tetra-scalar terms, all sitting at $\Lambda_4$.

The scalar and vector modes are each mixed at the quadratic level and the kinetic (and mass, in the case of the scalar) terms must be diagonalised in order to find the propagating modes \cite{Noller:2013yja}; doing so will then introduce an $N$ dependence to the previously $\mathcal{O}(1)$ coefficients in front of the $\Lambda_4$-suppressed terms, which then means that the actual cutoff can in fact be much lower.

Grouping the scalars into a column vector $\pi$, we can write their kinetic terms as $\mathcal{L}_{(\partial \pi)^2} \propto \pi^{\mathrm{T},\mu} K \pi_{,\mu}$. To find the propagating modes we must diagonalise the `kinetic matrix', $K$, and then canonically normalise by dividing each mode by the square root of the appropriate eigenvalue of $K$; an analogous proceedure applies for the vectors.\footnote{For the scalars we must also diagonlaise their mass matrix, however this turns out not to affect the scaling with $N$.}

Before introducing the plaquette, the kinetic terms involving $\pi_{(N,1)}$ are
\be
\pi_{(N,1)}^{,\mu} \pi_{(N,1),\mu} - \pi_{(N,1)}^{,\mu} \left( \pi_{(1,2),\mu} + \pi_{(N-1,N),\mu} \right),
 \ee
which, upon the plaquette substitution (just taken to lowest order, $\pi_{(N,1)} = - \sum_i \pi_{(i,i+1)}$, since here we are only interested in overall quadratic terms), becomes
\be
\sum_{i,j} \pi_{(i,i+1)}^{,\mu} \pi_{(j,j+1),\mu} + \sum_i \pi_{(i,i+1)}^{,\mu} \left( \pi_{(1,2),\mu} + \pi_{(N-1,N),\mu} \right).
\ee
Thus the kinetic matrix takes the form
\be
K = 
\begin{pmatrix}
2 & -1 & \\
-1 & 2 & \ddots \\
& \ddots & \ddots
\end{pmatrix}
+
\begin{pmatrix}
2 & 2 & \cdots \\
2 & 2 & \cdots \\
\vdots & \vdots & \ddots
\end{pmatrix}
+
\begin{pmatrix}
1 &\cdots & 1 \\
 & & \\
1& \cdots & 1
\end{pmatrix}
+
\begin{pmatrix}
1 & & 1 \\
\vdots & & \vdots \\
1&  & 1
\end{pmatrix}, \label{scalar kinetic matrix}
\ee
where blank entries are zero and the ellipsis denotes repetition, so the first matrix is tri-diagonal, the last only has non-zero entries in the first and last columns etc.
The first term is just the kinetic matrix for a line graph of length $N$ (see figure \ref{fig-graph example} (c)). Upon diagonalisation and normalisation \eqref{larger plaquette A} becomes
\be
\sum_{i=1}^{N-2} \sum_{j=i+1}^{N-1} \pi^{,\mu}_{(i,i+1),\lambda} \pi^{,\lambda}_{(j,j+1)} \propto \sum_{n,m=1}^{N-1} \left( \frac{1}{\sqrt{\lambda_n \lambda_m}} \sum_{i=1}^{N-2} \sum_{j=i+1}^{N-1} (v_n)_i (v_m)_j \right) \tilde{\pi}_{n,\lambda}^{,\mu} \tilde{\pi}_m^{,\lambda}, \label{A plaquette largest term}
\ee
where $v_n$ is the $n^{\text{th}}$ normalised eigenvector of $K$, $\lambda_n$ the corresponding eigenvalue, and $\tilde{\pi}$ the propagating modes. Taking this sum to be dominated by terms involving the smallest eigenvalue of $K$, which numerical investigations reveal to decrease to zero as $N^{-2}$, and taking $(v_n)_i \sim N^{-\frac{1}{2}}$, we find that the largest coefficent in \eqref{A plaquette largest term} scales like
\be
 N^2 \sum_{i=1}^{N-2} \sum_{j=i+1}^{N-1}\left( \frac{1}{\sqrt{N}} \right)^2 \sim N^3.
\ee
Remarkably the validity of these simple arguments is borne out by full numerical analysis of \eqref{A plaquette largest term}; we should also diagonalise the mass matrix for the scalar fields as well, however this turns out not to affect the scaling with $N$.

In the absence of a cycle the vectors are not mixed at quadratic level, which however is changed by the presence of a cycle and introduction of a plaquette:
\be
\partial_{[\mu} A_{(N,1)\nu]} \partial^{[\mu} A_{(N,1)}^{\nu]} \to \sum_{i,j} \partial_{[\mu} A_{(i,i+1)\nu]} \partial^{[\mu} A_{(j,j+1)}^{\nu]}
 \ee
Thus the kinetic matrix for the vectors takes the form
\be
K = 
\begin{pmatrix}
1 &  & \\
 & \ddots & \\
& & 
\end{pmatrix}
+
\begin{pmatrix}
1 & 1 & \cdots \\
1 & 1 & \cdots \\
\vdots & \vdots & \ddots
\end{pmatrix}, \label{vector kinetic matrix}
\ee
and we see that $\sum_{i=1}^{N-1} A_{(i,i+1)}$, which is precisely the combination appearing in the $\frac{1}{\Lambda_4^4} \partial A (\partial^2 \pi)^2$ terms, is an eigenvector, with eigenvalue $N$.

Therfore schematically we find
\be
\frac{1}{\Lambda_4^4} \partial A (\partial^2 \pi)^2 \sim \frac{N^\frac{5}{2}}{\Lambda_4^4} \partial \tilde{A} (\partial^2 \tilde{\pi})^2 \qquad \text{and} \qquad \frac{1}{\Lambda_4^8} (\partial^2\pi)^4 \sim \frac{N^6}{\Lambda_4^8} (\partial^2 \tilde{\pi})^4,
\ee
for the terms with the largest coefficients, and where a tilde indicates a propagating mode. Thus the cutoff \emph{decreases} as $\Lambda \sim N^{-5/8}$ for the first terms and $\Lambda \sim N^{-3/4}$ for the latter, which is interesting as it is marginally quicker than if one looks just at the $\Lambda_3$ suppressed terms for which one finds $\Lambda \sim N^{-1/2}$ \cite{deRham:2013awa, Noller:2013yja}.

The fact that the strong coupling scale decreases as the number of sites is increased is what prevents one from taking the continuum limit of the circle theory and arriving at Einstein gravity compactified on a circle. And we now see that the problem is even more severe if one formulates the theory in terms of metrics, rather than vierbeine.

It is worth now making contact with other work that has been done linking dimensional deconstruction and multi-gravity. In \cite{deRham:2013awa} it is noted that taking higher dimensional GR and naively discretising the metric in the dimension to be compactified will involve interaction terms polynomial in $g^{(i+1)}_{\mu\nu} - g^{(i)}_{\mu\nu}$ (where $g^{(i)}_{\mu\nu}$ is the effective lower dimensional metric at position $i$ in the discretised dimension), which will necessarily introduce a Boulware-Deser ghost \cite{Boulware:1973my}. We have now shown that even when one uses interactions which are individually ghost-free, constructing an extra gravitational dimension using metrics will introduce a ghost (in essence ours is a `bottom up' approach).

Secondly, in \cite{ArkaniHamed:2001ca, ArkaniHamed:2002sp, ArkaniHamed:2003vb, Schwartz:2003vj} it is argued that the truncated KK theory corresponds not to a single cyclic theory graph, but to a \emph{complete} graph,\footnote{i.e. one in which every site is linked to every other} in which the interaction strength for a given link decays as a power law in the distance in the extra dimension between the two sites (and thus the theory is non-local in the extra dimension). Given our results it would be interesting to see if, and how, when built using metrics, such a construction could lead to a cancellation of the terms sitting below $\Lambda_3$.

\section{Conclusions} \label{sec-conc}
        
In this paper we have answered a key question which remained concerning the consistency of theories of multiple, interacting spin-2 fields: does a cycle of interactions, when formulated using metrics, lead to the presence of a ghost (which is not present in the absence of the cycle)? We have shown that, even when the individual interaction terms are ghost-free, with a strong coupling scale of $\Lambda_3 = (m^2 M_\text{Pl})^{1/3}$, the cycle introduces higher-derivative terms, suppressed by the lower scale $\Lambda_4 = (m^3 M_\text{Pl})^{1/4}$, which will inevitably lead to the appearance of a ghost associated with that scale.

This was demonstrated in two ways: i) by using a plaquette construction to eliminate the `extra' \St field which is introduced due to the number of symmetry breaking interactions being larger than the number of broken symmetries, and ii) by introducing a reduced number of \St fields at the start. Both methods give the same form for the dangerous terms which appear, confirming the validity of this result.
We have also investigated the structure of interactions in the vielbein version of the theory and argued why the same ghost does not appear, which it should not, since this version is known to be ghost free \cite{Hinterbichler:2012cn}.

This result is interesting not just intrinsically, but also for its relation to dimensional deconstruction; the consequences on the latter of a ghost in the metric version we have examined by considering cycles of general size $N$ and we find that the previously noted problem of a low strong coupling scale which decreases like $N^{1/2}$ \cite{deRham:2013awa} is even more pronounced when the ghost is taken into account. More specifically the cutoff of the theory will in this case decrease like $N^{3/4}$, a further impediment to taking the continuum limit (and recovering the full KK theory), at least in the metric version.

Further work remains to be done exploring the link between cyclic theories and dimensional deconstruction, and it would be especially interesting to see if and how the ghosts which we have found here are present when one directly truncates the full KK theory. Similarly it would be worthwhile and useful to investigate whether it is possible to remove the ghosts via a suitable combination of cycles, and especially whether a complete graph with interaction strengths which decay with distance in the compactified dimension, such as described in \cite{Schwartz:2003vj}, would lead to the cutoff of the theory returning to $\Lambda_3$.
\\

\noindent {\bf Acknowledgements: } JHCS is supported by STFC. JN acknowledges support from STFC, BIPAC and the Royal Commission for the Exhibition of 1851. PGF was supported by STFC, BIPAC and the Oxford Martin School.

\appendix

\section{Necessity of plaquettes}\label{necessity of plaquettes}

In this appendix we will investigate the consequences of not eliminating one of the \St fields via construction of a plaquette; for concreteness and simplicity we will work in $D = 4$, with a trimetric theory in which all the interaction terms are the second symmetric polynomial $e_2 \left( \sqrt{g^{-1}f} \right) = \frac{1}{2}\left( \mathrm{tr}\sqrt{g^{-1}f}^2 - \mathrm{tr}g^{-1}f \right)$, and all Planck masses and interaction strengths are equal.

After introducing the \St fields and expanding about a flat background
\be
g_{(i)\mn} = \eta_{\mn} + h_{(i)\mn}, \qquad Y^\mu_{(i,j)} = x^\mu + A^\mu_{(i,j)} + \partial^\mu \pi_{(i,j)},
\ee
the scalar-tensor interaction terms which will survive in the decoupling limit are
\be
\mathcal{L}_{h\pi} = \sum_{i = 1}^3 \sum_{n=0}^{2} \hat{\alpha}_n h_{(i) \mn} \left( X^{\mn}_{(n)}(\pi_{(i,i+1)}) + X^{\mn}_{(n)}(\phi_{(i-1,i)}) \right),
\ee
where $\phi_{(i,j)}$ is the dual galileon field associated with $\pi_{(i,j)}$, $X^{\mn}_{(n)}(\pi)$ are transverse tensors involving $n$ factors of $\partial^2 \pi$, and $\hat{\alpha}_n = \frac{(3-n)!}{n! (2-n)!} $. (See appendix \ref{dual fields} and \cite{Noller:2013yja} for more details.)

We then perform a linearised conformal transformation to remove the scalar-tensor mixing at quadratic order
\be
h_{(i)\mn} \to h_{(i)\mn} - \frac{1}{2} \alpha_1 \left( \pi_{(i,i+1)} + \phi_{(i-1,i)} \right) \eta_{\mn},
\ee
which leads to the pure scalar part of the action
\be
\mathcal{L}_{\pi} = - \frac{1}{2} \hat{\alpha}_1 \sum_{i,n} \alpha_n \left( \pi_{(i,i+1)} + \phi_{(i-1,i)} \right) \left( \Ltd_{(n)}(\pi_{(i,i+1)}) + \Ltd_{(n)}(\phi_{(i-1,i)}) \right),
\ee
where $\alpha_n = (1 - \frac{1}{2} \delta_{n,1} ) (4 - n) \alpha_n$, and $\mathcal{L}^\mathrm{TD}_{(n)}(\pi)$ is the total derivative combination of $n$ copies of $\partial^2 \pi$. Finally we re-express terms involving the dual fields $\phi$ in terms of the $\pi$ fields
\begin{align}
\sum_n \alpha_n \pi_i \Ltd_{(n)}(\phi_j) &= \sum_n \beta_n \pi_i \Ltd_{(n)}(\pi_j), \\
\sum_n \alpha_n \phi_i \Ltd_{(n)}(\pi_j) &= \sum_n \gamma_n \pi_i \Ltd_{(n)}(\pi_j), \\
\sum_n \alpha_n \phi_i \Ltd_{(n)}(\phi_j) &= \sum_n \delta_n \pi_i \Ltd_{(n)}(\pi_j),
\end{align}
see \cite{multi-gal_from_multi-grav} for more details. Having $\beta_1 = 
\gamma_1 = - \alpha_1$ and $\delta_1 = \alpha_1 + \frac{1}{2} \alpha_0$, the Lagrangian is then
diagonalised at the quadratic level by the modes
\begin{align}
\chi_1 &= -\frac{1}{\sqrt{2}} \left( \pi_{(1,2)} - \pi_{(3,1)} \right) \\ 
\chi_2 &= -\frac{1}{\sqrt{6}} \left( \pi_{(1,2)} - 2 \pi_{(2,3)} + \pi_{(3,1)} \right) \\ 
\chi_3 &= \frac{1}{\sqrt{3}} \left( \pi_{(1,2)} + \pi_{(2,3)} + \pi_{(3,1)} \right),
\end{align}
of which the third is special since its eigenvalue of the kinetic and mass matrices is zero, and hence it drops out of the action at quadratic order! This is an indication that it might be acceptable to not use a plaquette since an appropriate combination corresponding to the constraint (\ref{constraint}) will then drop out leaving just two propagating modes. However this means that $\chi_3$ must not reappear other than linearly in higher order interaction terms. (It may appear linearly since in that case partial integration allows us to remove all the derivatives acting on $\chi_3$, reducing its role to a Lagrange multiplier enforcing a constraint on the dynamics of $\chi_1$ and $\chi_2$.) In particular, if it does appear as more than a Lagrange multiplier then, due to its lack of a kinetic term, it will be infinitely strongly coupled.

Looking first just at terms in which $\chi_3$ appears quadratically, table \ref{conditions} indicates the conditions that must be satisfied by $\alpha_n$, $\beta_n$, $\gamma_n$, $\delta_n$ for all of these terms to vanish.
\begin{table}[tp]
\centering
\begin{tabular}{ c r }
$\chi_2$ & $2(\alpha_2 + \delta_2) - (\beta_2 + \gamma_2) = 0$ \\
\hline
$\chi_2^2$ & $8(\alpha_3 + \delta_3) + 5(\beta_3 + \gamma_3) = 0$ \\
$\chi_1^2$ & $6(\alpha_3 + \delta_3) + (\beta_3 + \gamma_3) - 2 \beta_3 = 0$ \\
\hline
$\chi_2^3$ & $21(\alpha_4 + \delta_4) - \frac{3}{2}(\beta_4 + \gamma_4) = 0$ \\
$\chi_1 \chi_2^2$ & $\beta_4 - \gamma_4 = 0$ \\
$\chi_1^2 \chi_2$ & $16(\alpha_4 + \delta_4) + 3(\beta_4 + \gamma_4) - 4\beta_4 = 0$ \\
$\chi_1^3$ & $2(\alpha_4 + \delta_4) - (\beta_4 + \gamma_4) + 6\beta_4 = 0$
\end{tabular}
\caption{The row labelled $\chi_1^i \chi_2^j$ indicates the condition derived from the vanishing of the coefficient of the term $\chi_3 \chi_{3, \mn} \eta^{\mn \mu_1 \nu_1 \dots \mu_i \nu_i \rho_1 \lambda_1 \dots \rho_j \lambda_j} \chi_{1, \mu_1 \nu_1} \dots \chi_{1, \mu_i \nu_i} \chi_{2, \rho_1 \lambda_1} \dots \chi_{2, \rho_j \lambda_j}$; certain terms do not appear, e.g. $\chi_1 \chi_2$, since they vanish regardless of the values of $\alpha$, etc.} 
\label{conditions}
\end{table}
Although we have explicit expressions for $\alpha_n$, $\beta_n$ and $\delta_n$ \cite{multi-gal_from_multi-grav} we do not (yet) have an explicit expression for $\gamma_n$, however we can still check the consistency of the last two sets of conditions, which yield,
\begin{align}
\alpha_3 + \delta_3 &= \frac{5}{11}\beta_3,  \label{inf strong couple condition 1}\\
\alpha_4 + \delta_4 &= 0, \qquad \beta_4 = 0 = \gamma_4;  \label{inf strong couple condition 2}
\end{align}
whereas from the general expressions for an $e_m$ interaction term,
\begin{align}
\alpha_n &= (1 - \frac{1}{2} \delta_{n,1}) \frac{(D-n)!}{(D-m-1)! n! (D-n)!} \\
\beta_n &= (D-n)! \sum_{i=1}^{n} \frac{(-1)^i}{(D-i)!} \alpha_i \\
\delta_n &= -\frac{1}{(n+1)!} \sum_{i=0}^n \frac{(-1)^i (i+1)}{(n-i)!} \alpha_i,
\end{align}
we find for $D = 4$, $m = 2$:
\begin{align}
\alpha_3 = 0, \quad \delta_3 &= -\frac{1}{12}, \quad \beta_3 = 0 \\
\alpha_4 = 0, \quad \delta_4 &= -\frac{1}{120}, \quad \beta_4 = 0.
\end{align}
Thus the conditions (\ref{inf strong couple condition 1}) and (\ref{inf strong couple condition 2}) are not satisfied and $\chi_3$ will appear (at least) quadratically in the action, and thus will be infinitely strongly coupled hindering the analysis.

\section{Dual fields} \label{dual fields}

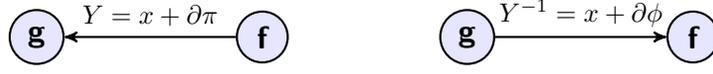
\begin{figure}[tp]
\centering
\begin{tikzpicture}[->,>=stealth',shorten >=0pt,auto,node distance=3cm,
  thick,main node/.style={circle,fill=blue!10,draw,font=\sffamily\large\bfseries},arrow line/.style={thick,-},barrow line/.style={thick,->},no node/.style={plain},rect node/.style={rectangle,fill=blue!10,draw,font=\sffamily\large\bfseries},sarrow line/.style={thick,->,shorten >=1pt},green node/.style={circle,fill=green!20,draw,font=\sffamily\large\bfseries},yellow node/.style={rectangle,fill=yellow!20,draw,font=\sffamily\large\bfseries}]
 
      \node[main node] (1) {g};
  \node[main node] (2) [right of=1] {f};
    \node[main node] (3) [right=2cm of 2] {g};
  \node[main node] (4) [right of=3] {f};

  \path[every node/.style={font=\sffamily\small}]
       (2) edge node [above] {$Y = x + \partial\pi$} (1)
          (3) edge node [above] {$Y^{-1} = x + \partial\phi$} (4.west);

\end{tikzpicture}
\caption{The galileon duality is equivalent to changing the direction of the \St link field.} \label{link direction swap}
\end{figure}

In this appendix we briefly review the galileon duality \cite{deRham:2013hsa} and its relation to multi-gravity theories \cite{Fasiello:2013woa,multi-gal_from_multi-grav}. The dRGT interaction terms possess a symmetry under interchange of the two metrics
\be
\sqrt{-g}\, e_m\left(\sqrt{g^{-1}(f \circ Y)}\right) = \sqrt{-(f \circ Y)}\, e_{D-m}\left(\sqrt{(f \circ Y)^{-1}g}\right),
\ee
and one can then gauge transform to get $\sqrt{-f}\, e_{D-m}\left(\sqrt{f^{-1} ( g \circ Y^{-1} ) } \right)$. This tells us that interactions of $f$ with the \St scalar $\pi$ are equal to those of $g$ with $\pi$, but with $m \to D-m$, and 
\be
\pi \to \phi \qquad \text{where} \qquad x + \partial \phi = (x + \partial \pi)^{-1}. \label{Galileon duality 1}
\ee
Since the interactions of $g$ with the \St scalar lead to a galileon Lagrangian for $\pi$, the interactions of $f$ will lead to a galileon Lagrangian for $\phi$. This is the essence of the \emph{galileon duality}: that the field redefinition (\ref{Galileon duality 1}) maps one galileon theory into another galileon theory, and is equivalent to changing the direction of the \St link field, as shown in figure \ref{link direction swap}.

The relation (\ref{Galileon duality 1}) can be extended to include the \St vector as well, in which case
\be
x + B + \partial \phi = x + \tilde{Z} = Y^{-1} = (x + Z)^{-1} = (x + A + \partial \pi)^{-1}.
\ee
Solving this, and disentangling the vector and scalar parts one finds
\begin{align}
\phi &= \sum_{n=1}^\infty \phi_n, \qquad \text{with} \qquad  \phi_n  = -\sum_{i=1}^{n-1} \frac{1}{i!} Z^{\nu_1} \dots Z^{\nu_i} \partial_{\nu_1 \dots \nu_i} \phi_{n-i},\\
B^\mu &= \sum_{n=1}^\infty B_n^\mu, \qquad \text{with} \qquad B^\mu_n = - \sum_{i=1}^{n-1} \frac{1}{i!} Z^{\nu_1} \dots Z^{\nu_i} \partial_{\nu_1 \dots \nu_i} B_{n-i}^\mu,
\end{align}
and inital values for the recursion relations
\begin{align}
\phi_1 &= -\pi, \quad \phi_2 = \frac{1}{2} \pi^{,\mu} \pi_{,\mu}, \\
B_1^\mu &= -A^\mu, \quad B_2^\mu = Z^\nu \partial_\nu A^\mu + A^\nu \partial_\nu^\mu \pi.
\end{align}

\section{Solution of equation (\ref{loop lst eom})} \label{loop lst eom solve}

Written in matrix notation, and ignoring terms which naively are suppressed by $\Lambda_2^2$, equation (\ref{loop lst eom}) becomes (note that $\omega$ is antisymmetric, whilst $A, B$, and $C$ are symmetric)
\be
\omega - ( \omega C + C \omega + A \omega B + B \omega A ) =  T, \label{syl-stein eq}
\ee
which is a combination of the Sylvester and Stein equations. It can be solved by use of the vectorisation operation (which turns an $n \times n$ matrix into a vector of length $n^2$, made of the concatenated columns of the matrix) and the identity
\be
\mathrm{vec} (XYZ) = ( Z^T \otimes X )\, \mathrm{vec} (Y),
\ee
where $\otimes$ represents the Kronecker product. Application of these to (\ref{syl-stein eq}) leads to
\be
\left( 1 - ( 1 \otimes C + C \otimes 1 + A \otimes B + B \otimes A ) \right) \mathrm{vec} (\omega) = \mathrm{vec} (T),
\ee
where 1 represents an identity matrix of the appropriate size. This can then be solved as system of linear equations, and in particular if the matrix on the left hand side is not singular (which we assume), we can multiply through by its inverse, which we then expand in a power series
\be
\left( 1 - ( 1 \otimes C + C \otimes 1 + A \otimes B + B \otimes A ) \right)^{-1} = \sum_{n=0}^\infty  ( 1 \otimes C + C \otimes 1 + A \otimes B + B \otimes A )^n,
\ee
and re-write
\be
1 \otimes C + C \otimes 1 + A \otimes B + B \otimes A = \frac{\partial}{\partial a} \frac{\partial}{\partial b} \frac{\partial}{\partial c} (a(A + cB) + b(1 + cC)) \otimes (a(A + cB) + b(1 + cC)) \Big|_{a = b = c = 0},
\ee
to get
\be
\mathrm{vec}(\omega) = \sum_{n = 0}^\infty  \frac{\partial^3}{\partial a_1 \partial b_1 \partial c_1} \dots \frac{\partial^3}{\partial a_n \partial b_n \partial c_n} (D_1 \dots D_n) \otimes (D_1 \dots D_n) \Big|_{a = b = c = 0} \mathrm{vec} (T),
\ee
where $D_i = (a_i(A + c_i B) + b_i (1 + c_i C))$. Finally turning each side back into a matrix gives 
\be
\omega = \sum_{n=0}^\infty \frac{\partial^3}{\partial a_1 \partial b_1 \partial c_1} \dots \frac{\partial^3}{\partial a_n \partial b_n \partial c_n} D_1 \dots D_n T D_n \dots D_1 \Big|_{a = b = c = 0}. \label{syl-stein eq sol}
\ee

\section{Terms in equation (\ref{loop lst eom})} \label{loop lst eom terms}

\begin{align}
T_1^{\mu\nu} =& -\left( \frac{1}{2} \tilde{X}^{\mu\lambda}_{(2,0)} + \frac{1}{18} \tilde{X}^{\mu\lambda}_{(0,3)} \right) G_{(2,1) \lambda}^{\nu} + \left( \frac{4}{3} \tilde{X}^{\mu\nu\lambda\rho}_{(1,0)} + \frac{1}{3} \tilde{X}^{\mu\nu\lambda\rho}_{(0,2)} \right) G_{{(2,1)} \lambda\rho} + \frac{\Lambda_2^2 }{18} \tilde{X}^{\mu\lambda}_{(0,3)} \Pi^\nu_{(2,1)\lambda} \\
A_1^{\mu\nu} =& \frac{1}{6} \Pi_{(2,3)}^{\mu\nu}, \qquad B_1^{\mu\nu} = \frac{1}{6} \Pi_{(2,3)}^{\mu\nu} \\
C_1^{\mu\nu} =& - \Pi_{(2,1)}^{\mu\nu} - \frac{2}{3} \Pi_{(2,3)}^{\mu\nu} - \frac{1}{72} (1 + \Pi_{(2,1)})^\mu_\lambda \left( 2 X^{\lambda\nu}_{(0,1)} + X^{\lambda\nu}_{(0,2)} + X^{\lambda\nu}_{(0,3)} + 18 X^{\lambda\nu}_{(1,0)} + 9 X^{\lambda\nu}_{(2,0)}\right) \nonumber \\
&- \frac{1}{3} (1 + \Pi_{(2,3)})^\mu_\lambda X^{\lambda\nu}_{(0,1)} + \frac{1}{12} \eta^{\mu\nu} \left( \mathcal{L}^\mathrm{TD}_{(1,0)} + 3 \mathcal{L}^\mathrm{TD}_{(0,1)} + \mathcal{L}^\mathrm{TD}_{(0,2)} \right) \\
T_3^{\mu\nu} =& \frac{1}{2}\left( \tilde{X}^{\mu\lambda}_{(1,2)} + \eta^{\mu\lambda} \right) G_{(2,3) \lambda}^{\nu} - 2 \tilde{X}^{\mu\nu\lambda\rho}_{(1,1)} G_{(2,3) \lambda\rho} + \frac{\Lambda_2^2}{2} \tilde{X}^{\mu\lambda}_{(1,2)} \Pi^\nu_{(2,3)\lambda} \\
A_3^{\mu\nu} =& -\Pi_{(2,1)}^{\mu\nu}, \qquad B_3^{\mu\nu} = \Pi_{(2,3)}^{\mu\nu}\\
C_3^{\mu\nu} =& - \Pi_{(2,1)}^{\mu\nu} - \frac{3}{4} \Pi_{(2,3)}^{\mu\nu} + \frac{1}{8} (1 + \Pi_{(2,3)})^\mu_\lambda \left( 2X^{\lambda\nu}_{(0,1)} + X^{\lambda\nu}_{(0,2)} + 8 X^{\lambda\nu}_{(1,0)} + X^{\lambda\nu}_{(1,1)} + X^{\lambda\nu}_{(1,2)} \right) \nonumber \\
&+ (1 + \Pi_{(2,1)})^\mu_\lambda X^{\lambda\nu}_{(0,1)} - \frac{1}{2} \eta^{\mu\nu} \left( 3 \mathcal{L}^\mathrm{TD}_{(0,1)} + 3 \mathcal{L}^\mathrm{TD}_{(1,0)} + \mathcal{L}^\mathrm{TD}_{(1,1)} \right).
\end{align}

\bibliographystyle{JHEP}
\bibliography{loops}

\end{document}